\documentclass[sn-basic,iicol]{sn-jnl_modified}

\jyear{2022}%

\begin{document}

\title[ ]{Exploring Descriptions of Movement Through Geovisual Analytics}


\author*[1]{\fnm{Scott} \sur{Pezanowski}}\email{scottpez@brightworldlabs.com}

\author[1]{\fnm{Prasenjit} \sur{Mitra}}\email{pum10@psu.edu}

\author[1,2]{\fnm{Alan M.} \sur{MacEachren}}\email{maceachren@psu.edu}

\affil*[1]{\orgdiv{Information Sciences and Technology}, \orgname{The Pennsylvania State University}, \orgaddress{\street{Westgate Building}, \city{University Park}, \postcode{16802}, \state{Pennsylvania}, \country{United States of America}}}

\affil[2]{\orgdiv{Department of Geography}, \orgname{The Pennsylvania State University}, \orgaddress{\street{Walker Building}, \city{University Park}, \postcode{16802}, \state{Pennsylvania}, \country{United States of America}}}


\abstract{Sensemaking using automatically extracted information from text is a challenging problem. In this paper, we address a specific type of information extraction, namely extracting information related to descriptions of movement. Aggregating and understanding information related to descriptions of movement and lack of movement specified in text can lead to an improved understanding and sensemaking of movement phenomena of various types, e.g., migration of people and animals, impediments to travel due to COVID-19, etc. We present GeoMovement, a system that is based on combining machine learning and rule-based extraction of movement-related information with state-of-the-art visualization techniques. Along with the depiction of movement, our tool can extract and present a lack of movement. Very little prior work exists on automatically extracting descriptions of movement, especially negation and movement. Apart from addressing these, GeoMovement also provides a novel integrated framework for combining these extraction modules with visualization. We include two systematic case studies of GeoMovement that show how humans can derive meaningful geographic movement information. GeoMovement can complement precise movement data, e.g., obtained using sensors, or be used by itself when precise data is unavailable.}




\maketitle

\section{Introduction}\label{sec1}

Automated methods proposed by the natural language processing and information retrieval communities often form the basic building blocks in an application. However, in this paper, we argue that such automated tools, even though they have achieved some level of maturity, are not enough for the needs of the end-users especially for
domains that require higher-level information assimilation and cognition like foraging and sensemaking over spatial information. For example,~\citet{lai2022nlp} have recently used natural language processing (NLP) to understand context in the extraction and geocoding of historical floods, storms, and adaptation measures. They extend the state-of-the-art for low-level information extraction, e.g., named entity extraction and geocoding, but do not provide a holistic understanding of the story underlying these events.

We posit that our research community needs to ``see the forest for the trees." Sensemaking is an integral part of
information processing, and tighter coupling between the lower levels (information extraction) and the higher levels (information understanding and sensemaking) can improve the state-of-the-art. Specifically, we call for the community to look more at ``higher-level tools and systems" that enable end-users to complete tasks. Towards this goal, we study the case of extracting geospatial information from text using visual analytics (VA)~\cite{andrienko2020va,yuan2021va} to perform tasks over the extracted data.

Since text is unstructured data and the information within the text is often messy, the output from computational techniques includes associated errors and is not sufficient to explore mentions of movement in text without human expertise. VA can address this issue through human-in-the-loop strategies that enable analysts to work iteratively with computational methods that extract knowledge from messy data, cope with uncertainties in computational results, and improve those results over time~\citep{endert2014viz,robinson2017geoviz}. VA is especially suitable for big, diverse, messy data that can be interpreted differently~\citep{tapia2021spatio,angelini2018visual,ninkov2019vaccine,snyder2020,MacEachren2011a}.

Our research objective was to determine if computational techniques and geovisual analytics can leverage \textbf{large volumes of movement statements to enable an end-user to understand the movement described} quickly. If successful, research can then take advantage of the wealth of movement data found in written descriptions about people, wildlife, goods, and other things moving throughout our world. Text statements about movement can be used to understand what is moving, when it is moving, why it is moving, and how it is moving.

For our research, ``geographic movement" refers to the movement of people, animals, objects, goods, information, natural physical processes, and similar things through spaces ranging in size from multiple buildings to the whole earth. We applied computational methods to identify and extract movement statements,
and present them in GeoMovement, a human-in-the-loop
web-based geovisual analytics system for  identification, processing, and exploration of descriptions of movement. 
GeoMovement involves computational 1) cleaning of the messy text, 2) predicting the statements that describe movement using a machine learning (ML) model, 3) applying Geographic Information Retrieval (GIR) techniques to identify places mentioned, and 4) predicting statements that describe restricted movement or desired movement that is not possible (hereafter referred to as ``impaired movement"). While there is substantial research on some of these subtasks, \textbf{integrating these techniques with VA} and demonstrating its success in our chosen domain is the main contribution of this paper.

While some progress has happened in processing descriptions of movement in text, there is very little work on detecting and understanding descriptions of impaired movement. For example, the COVID-19 pandemic prompted us to focus on impaired movement due to the importance of disruptions and restrictions in global movement patterns of people and, perhaps equally important, the movement of goods like food and medicines. Therefore, another important contribution of this work is that we show an adaptation of an existing approach developed to detect negation in picture descriptions~\citep{van2016pragmatic} can successfully uncover impaired geographic movement in text documents.

Specifically, we integrated (and adapted or extended) many existing computational and VA methods to produce a system that supports information foraging related to geographic movement as reflected in text statements. GeoMovement is unique in identifying movement statements and filtering them by place and time.

\begin{figure*}[htb]%
	\centering
	\includegraphics[width=\linewidth]{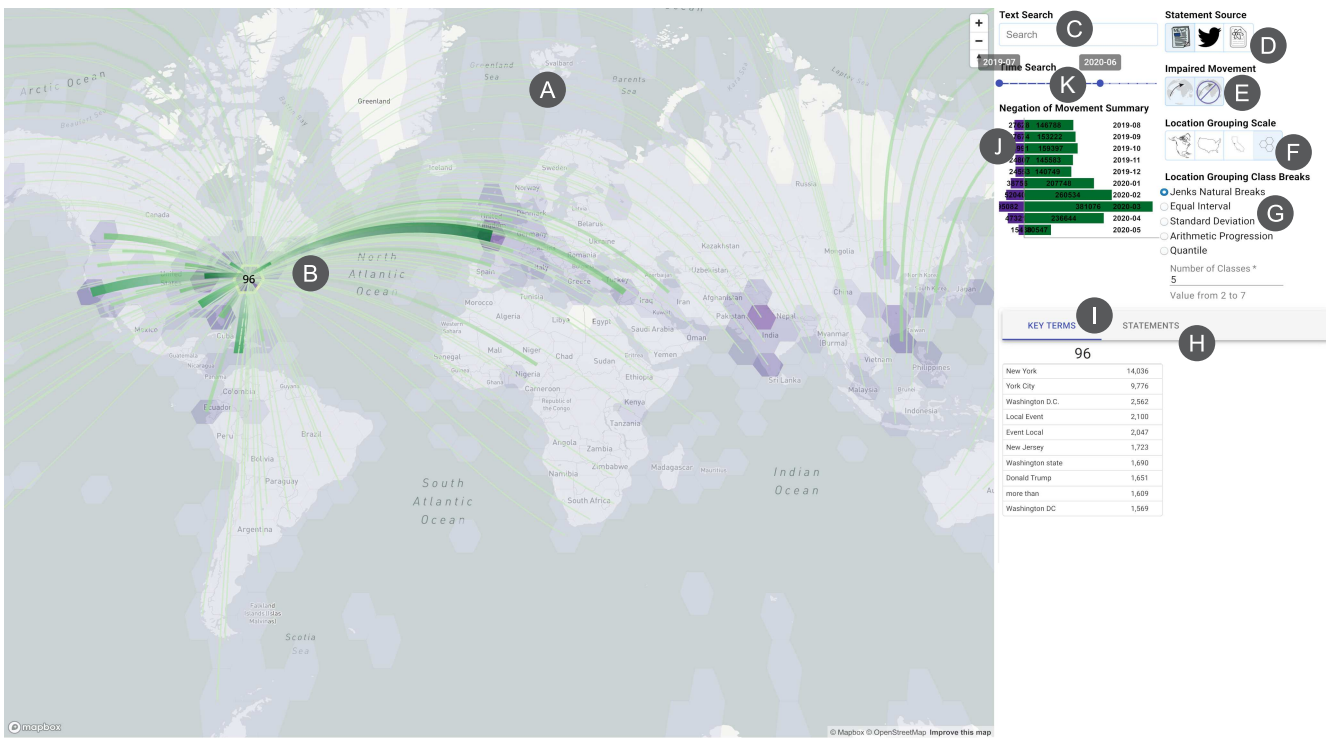}
	\caption{The GeoMovement user interface: A) Main mapping interface with hexagons selected for place aggregation B) Connection lines between co-occurring place mentions from an area near New York City selected C) Free text key term search D) Text data source buttons E) Normal movement vs. impaired movement buttons F) Geo-binning buttons G) Classification method for bins and connection lines colors and the number of classes chooser H) Statements that match the search I) Most common bi-grams that match the search J) Two-sided Temporal Bar Chart that shows the number of statements, by date, with a divided bar depicting normal movement on the right (green) and negated movement on the left (purple) K) Time range slider.}
	\label{fig:fig1}
\end{figure*}

Figure~\ref{fig:fig1} shows GeoMovement's user interface. Users can search and filter based on search terms, the statements' dates, and impaired movement status. The statements originate from three sources ingested into GeoMovement to demonstrate its capabilities and utility for investigating movement:

\begin{itemize}
	\item 398 thousand News articles from August 2019 to May 2020,
	\item 328 million Twitter tweets from February 2020 to May 2020,
	\item 15.6 thousand Scientific articles from August 2019 to November 2020.
\end{itemize}

\noindent
Over 520 million total statements contain diverse movement patterns, things moving, geographic coverage, and temporal differences.

Existing geographic movement research has improved analysis methods~\citep{dodge2012similarity,dodge2016future,dodge2016movement,dodge2016traj,graser2021movement,graser2020massive,graser2020trajectory,soares2017,huang2017} and shown how these methods can derive valuable information about human movement and wildlife movement~\citep{wang2020trajectories,dodge2014vultures,miller2019movement,zhu2021travel,li2021trans}. However, partly due to the challenges and because computational techniques to address them are relatively new, this prior research focused on geographic movement in precise movement trajectories from sensors such as GPS and mostly ignored movement described in text documents. The limited existing research to analyze movement described in text has only focused on narrow tasks like mapping route descriptions.
This work fills a gap in the research to support the study of geographic scale movement by developing and demonstrating methods for analyzing movement described in text documents at a broad scale. 
We anticipate that our effort will spur more research to leverage this under-utilized geospatial movement information in text documents.

We present two case studies (and a third case study in a supplementary video demonstration) related to the global pandemic that demonstrate the potential of our approach and shows, despite the messy data and imprecise computational predictions, a human can derive meaningful and essential information about geographic movement described in text with GeoMovement. The case studies also demonstrate that GeoMovement can support multi-scale information foraging through vast volumes of messy text statements about geographic movement. Furthermore, the space-time concept/attribute filtering methods implemented effectively narrow in on information relevant to an analyst's objectives.

\section{Related Work}\label{sec:sec2}

First, we review related efforts to analyze various geographic movement types described in text documents. Existing research into studying descriptions of geographic movement has primarily focused on reconstructing route descriptions. These routes (driving directions~\citep{jaiswal2012geocam,drymonas2010routes}, hiking and general route descriptions~\citep{moncla2014hiking,moncla2014itinerary,moncla2016itinerary,Piotrowski2010a}, and historical exploration routes~\citep{bekele2016historic,blank2015historic}) form a constrained subset of movement statements that simplify and thus do not address many
of the
challenges with a broader set of movement statements. Other research took the opposite approach of textualizing (convert routes to description) to take advantage of the benefits of text~\citep{chu2014taxi,aldohuki2017taxi}. 
Additionally,~\citet{huang2020nlp} showed how geovisual analytics could improve the retrieval of trajectories in a search. But, their focus on text analytics was only on users' queries of the data (which was precise sensor-based trajectories). Furthermore, in addition to addressing narrow domains, most of this research has restricted the data to a small amount of text.

The GeoCAM project~\citep{jaiswal2012geocam} created an application that identified, extracted, and generated maps of route directions found on webpages providing textual (often formatted) directions to reach a location. While the GeoCAM project complements the research we present here, their problem is
simpler since route directions 
are just a small subset of movement descriptions. Furthermore, route directions typically follow a semi-structured pattern that allows for simpler ML models and rules-based approaches. In a precursor to the GeoCAM project, \citet{drymonas2010routes} used similar techniques to map route directions. These projects encouraged future work like ours to go beyond route directions to general movement descriptions~\citep{klippel2008linguists}.

Second, we describe related efforts that use geovisual analytics on geographic movement described in text documents. The complicated nature of analyzing place in text documents, especially the need to represent spatial relationships best rendered visually on a map, has prompted other researchers to take a geovisual analytics approach. SensePlace was a system to analyze place mentions in text and pull information from other sources to aid analysis~\citep{Tomaszewski2011}.
SensePlace2 developed and applied geovisual analytics to methods that focus on the extent to which Twitter users' tweet location compared to the places they discuss in their tweets~\citep{MacEachren2011a}. \citet{Robinson2013b} performed a user study with experts that showed both the advantages, usefulness, and difficulties of such a system for crisis management. SensePlace3 extended this effort by advancing the geovisual analytics techniques and scaling the system to work on millions of tweets per month, thereby improving analysis~\citep{Pezanowski2017}.

The SMART system complements the SensePlace versions' focus by providing a visual interface enabling human analysts to explore text's spatial, temporal, and topical components~\citep{snyder2020,karimzadeh2019geovisual}. SMART implemented advanced geovisual analytics techniques, including a tweet classifier to filter semantically and a cluster lens to visualize keywords at a large scale. However, like SensePlace3, their system did not focus on analyzing movement. Finally, the NewsStand system~\citep{teitler2008,samet2020} mapped places where news articles are written compared to the places they discuss but focuses less on geovisual analytics for analysis and more on correctly mapping the text.

A few other efforts use geovisual analytics and mapping systems to analyze places mentioned in tweets~\citep{Thom2012,Bosch2013,Bosch2011,felmlee2020women} and show the potential to take advantage of this geographic data source, albeit mainly focusing on tweets with a geocoded location that makes the challenge different. Mapping and geovisualizations have also been combined with topic analysis and network graph analysis to show similarities between cities~\citep{hu2017extracting}.

\citet{jamonnak2020videos} combined location information associated with videos and the narrations of those videos to show their locations on the map and the topics and sentiment discussed at those locations.
\citet{xu2018reviews} created a system to explore Yelp business reviews in areas and their change over time.
\citet{ma2020local} showed that geovisual analytics is vital to understanding
critical local places that need immediate help in a disaster from 911 call transcripts and clusters of specific crime types from police reports. Although these systems successfully demonstrated geovisual analytics on text, they focused on particular topics and did not consider movement. Therefore, they could not be applied to our goal of analyzing wide-ranging types of movement described in text. Since impaired movement detection is not the primary focus of this research, we describe work related to it in Section~\ref{sec:impaired}.

\section{Text Computational Processing Predictions}\label{sec3}

We first acquired three different sets of documents consisting of 398 thousand news articles, 328 million tweets, and 15.6 thousand scientific articles. Our document sources and the keyword and time parameters used to obtain them are described in detail in Appendix~\ref{sec:textsources}. We cleaned the documents using typical text pre-processing methods and applied computational techniques to identify places in the text, predict the statements that describe movement, and predict statements that describe impaired movement.

\subsection{Predicting Movement}\label{sec:predmovement}


In \citet{pezanowski2020geomovement}'s work, humans label the statements with a binary class as either describing geographic movement or not. Since they took this initial step and created a corpus to train a model, we can use this model to predict statements that describe movement. The prediction of this ML model is a probability value between zero (no movement) and one (movement). We set an arbitrary cut-off for GeoMovement to only show statements with a probability greater than 0.6 that the statements are about movement. This relatively loose threshold was arrived at by trial and error and is acceptable for all three sources.


\subsection{Predicting Geographic Location}

We used the GeoTxt system to perform GIR on our text sources~\citep{karimzadeh2013geotxt,karimzadeh2019geotxt}. We chose this because GeoTxt performs comparably or better than other state-of-the-art geoparsers and performs best without case sensitivity, which is common in Twitter data (one of our data sources)~\citep{gritta2018geoparsing}. 
A small evaluation corpus had an F1-score of 0.78 for geoparsing place names and an accuracy of 0.91 in resolving those place names correctly. Moreover, 
it performed even better on higher-order administrative places such as countries and states, which are common in our statements. For enterprise projects, paid commercial sources also exist from software companies such as Esri, Google, and Microsoft. We chose not to use these products because they do not allow for customization compared to GeoTxt, which is open-source software and
allows adjustments to the software in the future.

\subsection{Predicting Impaired Movement}\label{sec:impaired}

The global pandemic of 2020-2021 brought attention to global movement and how it spreads, and how the pandemic disrupted or prevented regular global movement. Because the pandemic highlighted the importance of analyzing disruptions to movement, we investigated potential strategies for detecting statements about impaired movement.

To detect impaired movement in our statements, we looked to adapt existing methods of negation detection in text. Negation detection strategies can potentially uncover statements about formal restrictions on movement (of the sort imposed by governments), decisions not to move taken by individuals for their safety, and impediments to movement created by limited public/commercial transport such as canceled flights due to the lack of passengers or ill crew.

In this section, first, we review existing related research on negation detection and its everyday use cases. 
Second, we describe how we adapted an existing approach from the literature that detects negation in picture descriptions~\citep{van2016pragmatic} to our challenge of detecting impaired geographic movement described in text. Third, we show how we improved upon our initial attempt to detect impaired movement using our geovisual analytics methods to analyze initial mistakes in predictions and then modify the rules specifically to detect impaired movement.

\subsubsection{Existing Approaches to Negation Detection}

Addressing negation in text has been identified as a problem in several existing works. For example, \citet{Fialho} had remarked that ``when a negation was involved in a sentence, the classifiers found more difficult to return the appropriate label" in the context of negation in sentences as identified as part of discourse representation structures. Negation detection in text is vital in challenges like automated summarization of medical reports~\citep{Vincze2008,slater2021negation}, summarizing picture descriptions~\citep{van2016pragmatic}, and as a hint in identifying sarcasm~\citep{reyes2014sarcasm}. \citet{hiremath2021sarcasm} show that sarcasm detection depends upon detecting negative sentences in positive situations and positive sentences in negative situations.

Much of the current state-of-the-art research on negation detection was influenced by a Workshop titled \textit{Resolving the Scope and Focus of Negation}~\citep{Morante2012}, which also produced labeled datasets that continue to be used in training and evaluating the success of new methods. Supervised ML-based solutions such as the LSM Network~\citep{zhao2021sentiment} have learned negative terms while performing sentiment mining automatically from large-scale training data. The current state-of-the-art method, NegBert, is based on ML~\citep{khandelwal-sawant-2020-negbert,khandelwal2020multitask}. Although this ML approach is the current state-of-the-art, the challenge in using ML approaches is the need for time-consuming labeling of large amounts of training data. NegBert was trained and tested on datasets designed explicitly for negation detection evaluation and therefore could not be used for our tangential challenge of detecting impaired movement. In the absence of training data, we show that rules-based approaches can still be used.

\citet{van2016pragmatic} and \citet{van2016building} have used rules to detect negation. They had humans annotate Flickr picture descriptions for negations and define categories of negations. This annotation exercise produced simple clues for negation, thereby allowing their rules-based approach to be effective on picture descriptions. In general, rules-based approaches have been proven to work in negation detection when the domain is relatively narrow and,
like most rule-based systems typically provides high precision but low recall. Rule-based methods can work fine in our application, where a sample of the negative sentences suffices,
but having false positives can result in incorrect conclusions.

\subsubsection{Applying Negation Detection to Descriptions of Geographic Movement}

Detecting impaired movement is similar to previous negation detection using specific key terms. However, what constitutes a negated word is ill-defined and varies
from domain to domain and problem to problem 
since terms, like \textit{canceled} or \textit{diverted}, would not always be considered negated. But, when applied to movement, they are. This ambiguity complicates the task. We investigated if we could adapt the current methods that are focused on the negation of words (ex. She does \textit{not} have cancer. The alarm clock did \textit{not} have the feature I wanted.) to detect impaired geographic movement (ex. Our flight to England was \textit{canceled}. The fruit was \textit{stuck} in Brazil because of initial fears early in the pandemic.). 

We used~\citet{van2016pragmatic}'s rules-based approach for negation detection. As discussed above, we adapted this rules-based approach, as opposed to an ML approach, to 1) avoid costly labeling of training data, 2) achieve transparency in how the results are obtained, and 3) because our needs in detecting impaired movement are relatively narrow, which the literature suggests~\citep{van2016pragmatic,van2016building,slater2021negation} is a good fit for a rules-based approach.

\citet{van2016pragmatic}'s method tags part-of-speech in text (such as verbs, nouns, adjectives, etc.) and then searches for a list of negation keywords, prefixes, and suffixes to detect negations. Some negation rules examples are a) exact negated words like ``no" and ``not," b) verbs that start with ``de," ``mis," ``dis," or c) adjectives that end with ``less." Overall, all rules are relatively simple and are easily understandable and reproducible. If the input sentence matches a rule, it containing negation.

We first applied \citet{van2016pragmatic}'s exact rules for negation to our statements previously predicted to describe movement (as described in Section~\ref{sec:predmovement}) to predict impaired movement. After 800,000 movement statements were predicted, we stopped and obtained a summary count. The model predicted about 28\% of these movement statements as describing impaired movement.

We selected a stratified random sample of 50 predicted impaired movement statements and 50 predicted normal movement statements. The initial model did not do very well to predict impaired movement correctly. There were 23 true positives where the statement was correctly predicted as impaired movement compared to 27 false positives where the statement was predicted to be impaired movement, but it was normal movement. It did slightly better in correctly predicting normal movement with 42 true negatives as opposed to eight false negatives. Table~\ref{tab:matrix} shows the confusion matrix for these predictions. These values equate to a precision of 0.46, a recall of 0.74, an F1-score of 0.57, and an accuracy of 0.65 on the stratified sample.

\begin{table}[htb]
\begin{center}
\caption{The confusion matrix for prediction of negated movement with unmodified rules from general negation.}\label{tab:matrix}%
\begin{tabular}{@{}|c|c|c|@{}}
\hline
 & Actual impaired  & Actual normal \\
\hline
Predicted impaired   & 23 (TP)   & 27 (FP)  \\ \hline
Predicted normal    & 8 (FN)   & 42 (TN)  \\
\hline
\end{tabular}
\end{center}
\end{table}

\subsubsection{Improving the Detection of Impaired Movement}

To improve our detection of impaired movement, we either added new rules or removed existing rules. Since there were more false positives than false negatives, this hints that rules should mostly be removed so that fewer statements are predicted to have impaired movement.

Not only did we analyze the errors in statement predictions, but we also consulted a list online of verbs common to movement see Appendix~\ref{sec:A} and \url{https://archiewahwah.wordpress.com/2019/04/16/movement-verbs-list/}). First, we removed rules specific to verbs that had the prefixes ``de," ``mis," and ``dis." These prefixes could rarely negate the verbs we found related to movement. Second, based on an online list of adjectives related to movement (see Appendix~\ref{sec:A} and \url{https://dspace.ut.ee/bitstream/handle/10062/18053/adjectives\_and\_adverbs\_of\_movement.html}), we removed the adjectives' rules beginning with ``a" and ``dis" because they are much too broad and did not make sense with those commonly related to movement. Third, we added the specific lemmas ``cancel," ``postpone," ``prevent," and ``avoid" because they relate to impaired movement. Lemmas allow us to match many different synonyms of these words that all mean the same.

These rules modifications improved predictions on the existing stratified random sample with a precision of 0.74, a recall of 0.76, an F1-score of 0.75, and an accuracy of 0.79. Therefore, the F1-score improved from 0.57 to 0.75.

Since we used the first set of 100 sampled statements to determine some of our rules, it would not be fair to rely on these improved metrics alone to conclude we improved our method. The rules may over-fit the sample. Therefore, we selected a second unseen test set of 100 statements using the same random stratified technique. On this unseen set, the original rules' prediction F1-score was 0.59, while our new rules were again much better with a 0.65 F1-score. Overall, our minor modifications to the rules produced substantial improvement. 

Finally, since we selected the two sets of 100 statements from new articles, as a final assessment, we sampled 100 statements from both the tweets and scientific articles using the same stratified random sampling technique. 
Predictions using our modified rules for impaired movement on the tweets resulted in an F1-score of 0.61 and an F1-score of 0.60 on scientific articles.
Therefore, our predictions for impaired movement performed slightly less accurately for these two sources than news articles, but still respectable. This lower accuracy is likely because both sources contain language more common for that audience (i.e., slang and other informal languages in tweets and technical language in scientific articles). Based on our experience modifying the rules for news articles, we estimate that additional rule changes would also improve predictions for tweets and scientific articles. However, based on our experience and the literature, we surmise that given resources to label a large amount of training data, using an ML approach as in~\citet{khandelwal-sawant-2020-negbert} would likely produce more accurate predictions. In summary, we are using a rules-based approach as a proof-of-concept prediction that is important to show the potential benefits of detecting impaired movement statements and use this as an attribute to analyze geographic movement described in text.

\section{Geovisual Analytics to Find Meaning in Descriptions of Movement}

GeoMovement is a web-based geovisual analytics system that allows users to explore descriptions of geographic movement. GeoMovement serves as an interface between the human and the data described above and is summarized in Figure~\ref{fig:fig1}. The statements' content, place mentions, date of creation, and impaired movement prediction are all searchable. The map visualizes place mentions in multiple levels of aggregate geo-bins. A geo-bin is many smaller places (e.g., cities) aggregated and displayed as one larger place that they all reside within (e.g., country). Geo-bins allow for a clear summary of location-based data.  By choosing particular places of interest on the map, the user can visualize the co-occurrences of places. The individual statements view completes the overview-first + detail approach. This workflow matches the information-seeking mantra of overview-first to gain an awareness of the information and details of interest on-demand~\citep{shneiderman1996eyes}. This section is divided into three subsections that follow the information seeking mantra that starts with the overviews, then options for the user to search to filter to statements of interest, then detailed views of the filtered statements.

For detailed information on GeoMovement, Appendix~\ref{sec:textsources} describes the sources and nature of the data sources and statements. In addition, Appendix~\ref{sec:B4} includes technical details of the application development that allow for fast user queries on large volumes of text, thereby enabling efficient sensemaking. An important technical component of our approach is our use of Elasticsearch (\url{https://www.elastic.co/})) as the primary information storage and retrieval software for GeoMovement. Elasticsearch is a search engine that accepts many search parameters like free text and time, and returns matching results ordered by most relevant to the user's search. In addition, it can group results by attributes like place mentions in the text. Most impressively, Elasticsearch does all of this and returns results very quickly, most often in the matter of milliseconds.

\subsection{Overviews}

There are two primary overview means to explore the movement descriptions. The first overview is the map (Figure~\ref{fig:fig1} at point A) that displays the number of statements spatially aggregated by their place mentions. Since we have a large number of statements from multiple sources, GeoTxt extracted over 98,000 unique place mentions from them. Displaying this large number of places on a map using points would likely be very confusing for users. Many places would overlap and quickly seeing overall patterns along and comparing quantities between places would be difficult. This is the primary reason for aggregating the statements place mentions. Coloring the polygon bins that contain places mentions a lot in the statements darker than those that are not mentioned frequently in the statements. Users get a clear overview of place mentions in the statements and a way to visually compare places.

We chose five shapes to spatially aggregate the number of statements by their place mentions. The geo-bins scales include continent, country, administrative 1 (the worldwide equivalent of a state in the United States), and two sizes of a hexagonal pattern. Each scale allows users to explore different types of movement data, like long-range bird migrations by continent, traded goods by country or administrative level, and detailed movement through the hexagons that do not adhere to political boundaries like the spread of disease. We chose hexagons as an aggregation shape since they tessellate and will distort values less than squares~\citep{birch2007hexagons,esri2021hexagons}. The user interface provides an option shown in Figure~\ref{fig:fig1} at point F where the user can choose the aggregate level.

The bin counts are divided into classes to ease visual comparisons between bins. The user has control of the bin count classification technique and number of classes, as shown in Figure~\ref{fig:fig1} at point G. The user has five choices on how to classify the aggregate counts. Jenks Natural Breaks, Equal Interval, Standard Deviation, Arithmetic Progression, and Quantile are options, and they can enter the number of classes between two to seven. These classification techniques are well accepted statistical methods to make the data more understandable.

We used the geostats JavaScript library (\url{https://github.com/simogeo/geostats}) to calculate the class breakpoints for the chosen users' classification method. Each classification method is valuable depending on the user search and resulting data. Darker colored bins represent areas that have a larger number of place mentions within the bin. The color scheme is a sequential color scheme chosen from ColorBrewer (\url{https://colorbrewer2.org/})~\citep{brewer2003color} to ensure the classes of statement counts are easily distinguishable.

The second overview in GeoMovement is a Two-sided Temporal Bar Chart. Research has shown that there are differences in how people understand geographic movement such as the those who think more spatially than others~\citep{liben1993space,ishikawa2016gis}. Based on this research, GeoMovement provides the user multiple views of the data. The Two-sided Temporal Bar Chart shows aggregate counts of statements grouped by the month they were published (Figure~\ref{fig:fig1} at point J). The oldest month is at the top and the youngest month is at the bottom. The Two-sided Temporal Bar Chart allows for visualization of the amount and changes over time, both in impaired movement on the left side and normal movement on the right side. When hovering the mouse over the chart, the number of statements that match the current user search are shown in bars in the foreground while the total number of statements in GeoMovement are shown in bars in the background. Figure~\ref{fig:fig6} shows an example of the number of statements that match the user search in the foreground once the user has hovered the cursor over the chart.

\subsection{Search}

After the overviews give the user an understanding of the data, they can begin filtering it through search options. The five key ways to search the data are 1) free-text search (Figure~\ref{fig:fig1} at point J), 2) buttons to select from the three sources for statements --- news articles, tweets, and scientific articles (Figure~\ref{fig:fig1} at point J), 3) buttons to select impaired movement statements or normal movement statements (Figure~\ref{fig:fig1} at point J), 4) a time-range slider (Figure~\ref{fig:fig1} at point J), 5) and clicking location(s) on the map to filter by location. Multiple features can be chosen by holding the Ctrl-key on their keyboard and selecting the next feature with a mouse-click. After any of the first four searches are performed, the geo-bins and Two-sided Temporal Bar Chart update to show counts of statements that match the searches. After the fifth search, the detailed views appear.

\subsection{Detailed Views}

The detailed views of statements and their attributes include a) connecting great circles drawn
on the map between place mentions in the statements, and co-occurring place mentions in the same statement, b) the five most common bi-grams for the set of statements matching the search, c) and a table showing the actual statements that match the search. All connections between places are aggregated to show places commonly used together. The aggregate count classes adhere to the users' currently chosen map classification method and total class number in a sequential green color scheme that is color-blind friendly, also selected from ColorBrewer~\citep{brewer2003color}.

It is important to note that the connection lines are completely accurate in showing movement between the places since they are drawn solely by the places' co-occurrence in a statement. For example, the statement below has three place mentions and therefore three possible movement pairs: Sydney–New York, Sydney–London, New York–London. The first two are correct concerning movement references in the statement, but the third is not correct because there is no direct movement between New York and London. One could argue that the movement could be from New York to Sydney and then next to London from Sydney, but this is not probable. However, these connection lines to give the user overall patterns of interest that they can confirm through inspection of the statements.

\begin{quote}
But this will be the first time a commercial flight is flown from Sydney to New York, and just the second time from Sydney to London, Qantas said~\citep{Garber2019}.
\end{quote}

Second, the ten most common bi-grams are shown for the set of statements that match the search and have place mentions in the chosen location bin. This provides details about the actual statements behind the overviews by showing the most common words and topics in the selected statements. Since many of the statements contain place mentions and the search location is an essential parameter for the matching statements, two-word place names are often in this bi-grams list. These place mentions may be valuable, but we also found that it was often more important to see bi-grams about topics and not necessarily places when exploring the data. Therefore, we allow users to double-click a bi-gram to remove the bi-gram from the list, and the next most common bi-gram appears, up to the 20th most common bi-gram. To re-populate the bi-grams list, the user can re-run the search. The user can also select multiple bins by holding the Ctrl-key and clicking another map bin. Users can then compare the most common bi-grams for both map bins to see differences between statements with place mentions in each bin.

To complete the overview first + detail approach to analysis, once the user found sets of statements of interest in the overviews and chose a map bin, the user can see the actual statements matching the search in a paged list. The displayed statements match all search parameters (when selecting multiple bins, statements can have place mentions in either bin). The user can scroll through the pages of statements. Each statement's published date is also shown. If there is a particular statement of interest, holding the Ctrl-key while clicking on the statement will open the original document on the Web in a new browser window.

GeoMovement's tight use of modern search engine technologies and Information Retrieval (IR) allows for extremely fast human-in-the-loop sensemaking for the most relevant information on movement. Again, Appendix~\ref{sec:B} provides further details, and our supplemental video showcases a real-time live demonstration of efficient knowledge discovery using GeoMovement.

\section{GeoMovement Assessment}\label{sec:assess}

First, we
show the challenge of interpreting movement described in statements without GeoMovement by discussing summary statistics of the data. Second, we present two case studies that show how our approach can retrieve information about movement from vast quantities of statements. A third case study is included in a supplemental video. It is recorded in real-time to show that a user can quickly extract meaningful information about movement despite the challenges 
posed by the large quantity of messy text, ambiguity in text descriptions, and imprecise computational predictions. 
Third, we show how these case studies also generated future GeoMovement needs. 
Finally, in Appendix~\ref{sec:C}, we discuss the skill level and hardware and software requirements for GeoMovement users.

\subsection{Illustrative Data Summary Statistics}

To show the value of GeoMovement, we created summary statistics of the data to clearly illustrate how it is unreasonable to think a human can analyze and understand large quantities of movement statements without such a system. We chose three keywords related to our case studies and three prominent place names: one being a country, one a state, and one a city. The number of our statements that match these keywords and place names is shown in Table~\ref{tab:counts}. To relate these statistics to our first case study in Section~\ref{sec:cases}, we filter GeoMovement's 520 million statements (36 thousand of those contain the term \textit{smuggling}, 275 thousand of them include the term \textit{gold} and 201 thousand mention \textit{India}) and efficiently identify important gold smuggling patterns around and in India. To produce these statistics, we used our GIR extracted place names to determine the number of statements that contain each selected place name. And, to find the number of statements containing each of the keywords, we searched the statements in a Postgresql database using a full-text search (\url{https://www.postgresql.org/docs/14/textsearch.html}) so that different variations of the same word will be matched (e.g. smuggling, smuggled, smuggle).

\begin{table*}[htb]
	\begin{center}
		\caption{Summary statistics of the number of statements that contain selected key terms and place mentions.}\label{tab:counts}%
		\begin{tabular}{@{}|c|c|c|c|c|c|c|@{}}
			\hline
			smuggling & gold & sports & London & California & India & statements \\
			\hline
			36 K & 275 K & 336 K & 589 K & 165 K & 201 K & 520 M  \\ \hline
		\end{tabular}
	\end{center}
\end{table*}

This exercise to generate summary statistics is meant to show that without such a geovisual analytics system like ours, it would be extremely difficult, if not unfeasible, to perform geographic, temporal, and attribute sensemaking of the described movement. Although GeoMovement’s 520 million statements are substantial, it is still a fraction of the accessible text available that could be included and analyzed in GeoMovement, given more development and computational resources.

\subsection{Case Studies}\label{sec:cases}

To confirm our claim that the geovisual analytics interface helps users understand and make sense from the statements and multiple computational predictions, we provide two detailed case studies below from different types of (prototypical, fictitious) potential users. The case studies presented provide evidence of usability. A third case study, given only in the video supplement (due to space limitations in the text), adds additional evidence about the flexibility and utility of GeoMovement to explore the mix of text data sources from different perspectives.

\subsubsection{Understanding International Crime Affecting India}

\begin{figure*}[htb]%
	\centering
	\includegraphics[width=\linewidth]{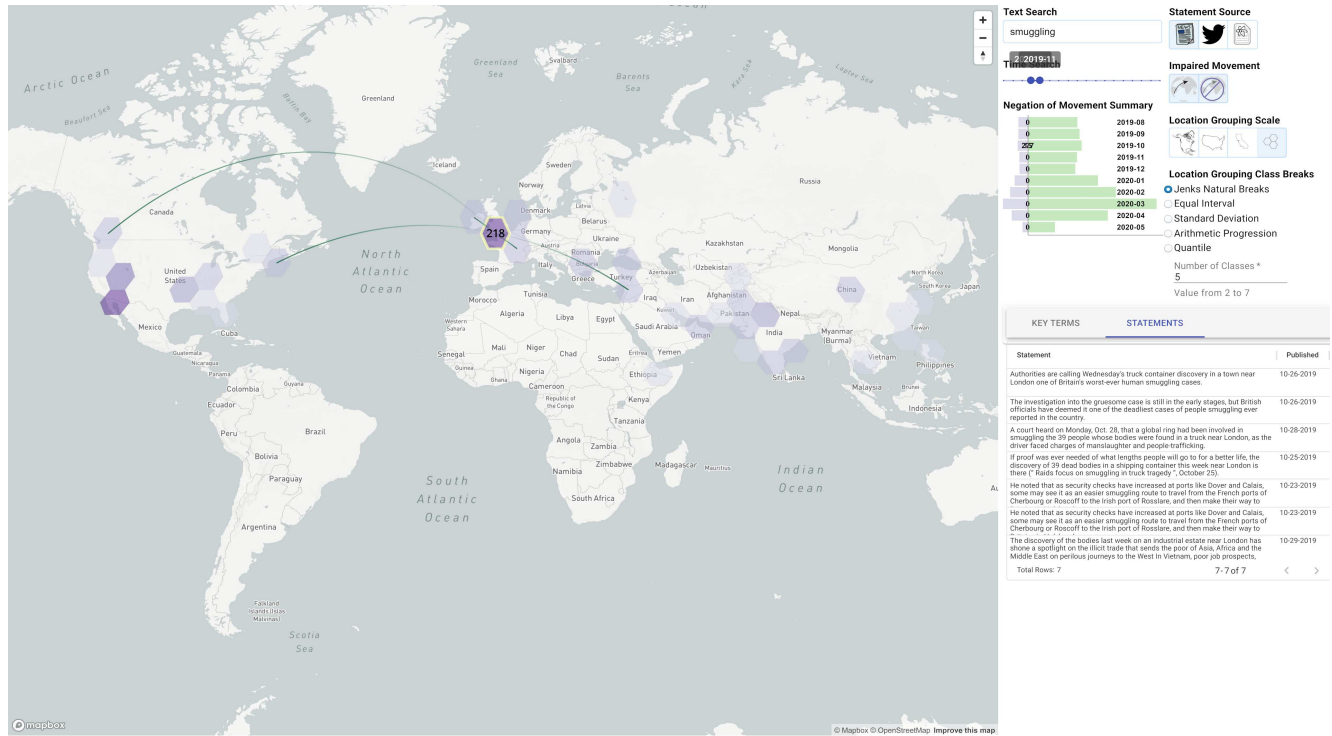}
	\caption{By filtering the search results for \textit{smuggling} to only show statements from October 2019, she sees many statements from a significant human trafficking event in England where many people lost their lives.}
	\label{fig:fig2}
\end{figure*}

Jennifer Lang is a college student who wants to write a class report about different types of international crime affecting India. She opens her web browser to GeoMovement and types \textit{smuggling} to begin her search. 
She notices that many statements involve smuggling are in October 2019, despite that month having fewer statements overall. She adjusts the time range slider to filter statements to that month and sees a hotspot of activity in England. She clicks the hexagon bin in England and is reminded of a significant human smuggling event in that month where many people lost their lives after being trapped in a truck that was smuggling them (Figure~\ref{fig:fig2}). Although this is a significant smuggling event from a British perspective and also highlighted in the U.S. news, she decides to look for other ways to focus on India.

\begin{figure*}[htb]%
	\centering
	\includegraphics[width=\linewidth]{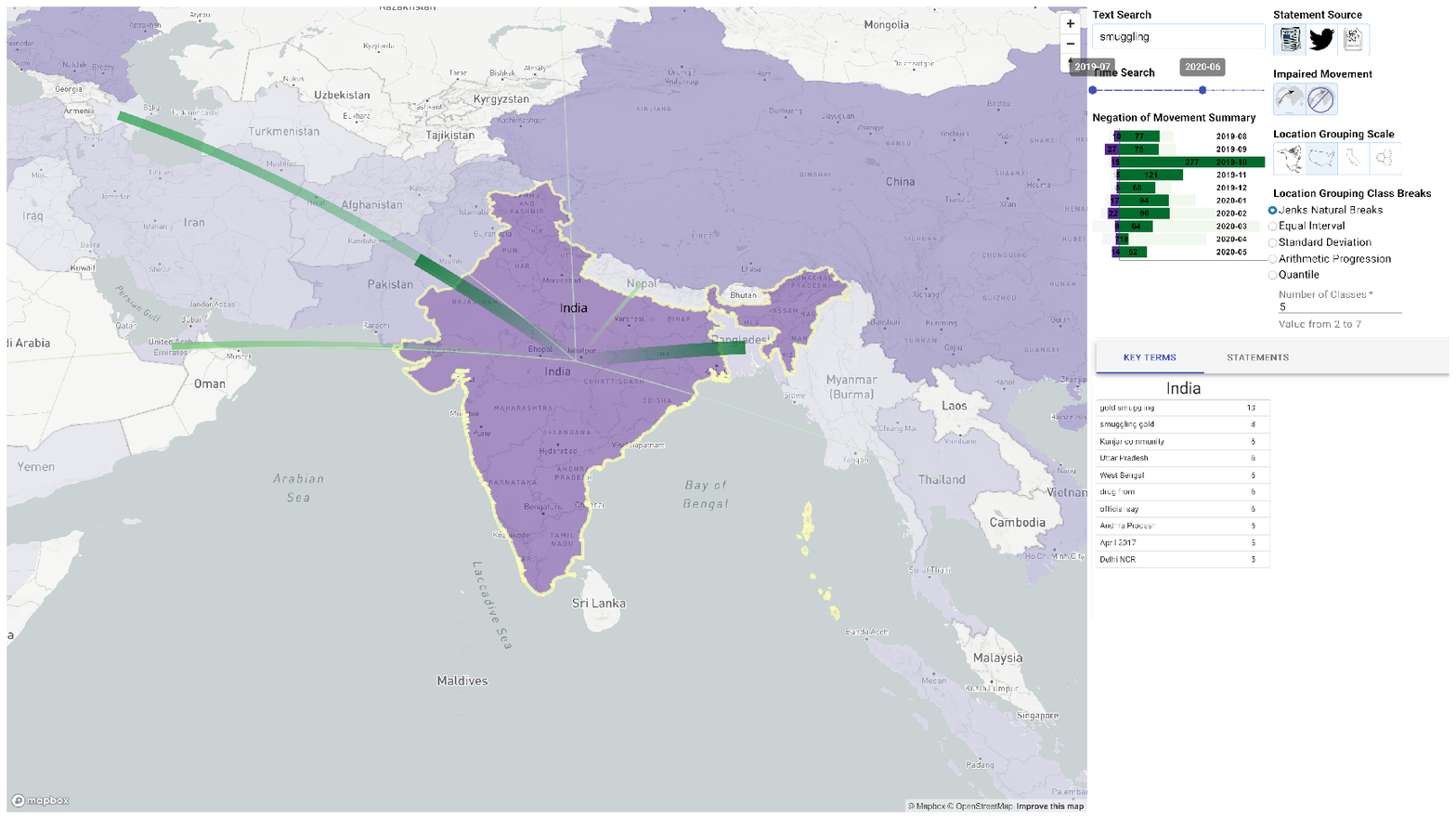}
	\caption{A search for \textit{smuggling} shows important location sources for India and a spike of activity in October 2009.}
	\label{fig:fig3}
\end{figure*}

As her next step, she chooses to aggregate place mentions by the country level and select India. Figure~\ref{fig:fig3} shows the results, and she sees that most results related to smuggling are affecting India from the neighboring countries of Pakistan and Bangladesh, and a few more countries. In the bigrams list, she sees many references to drugs and gold being smuggled into India.

\begin{figure*}[htb]%
	\centering
	\includegraphics[width=\linewidth]{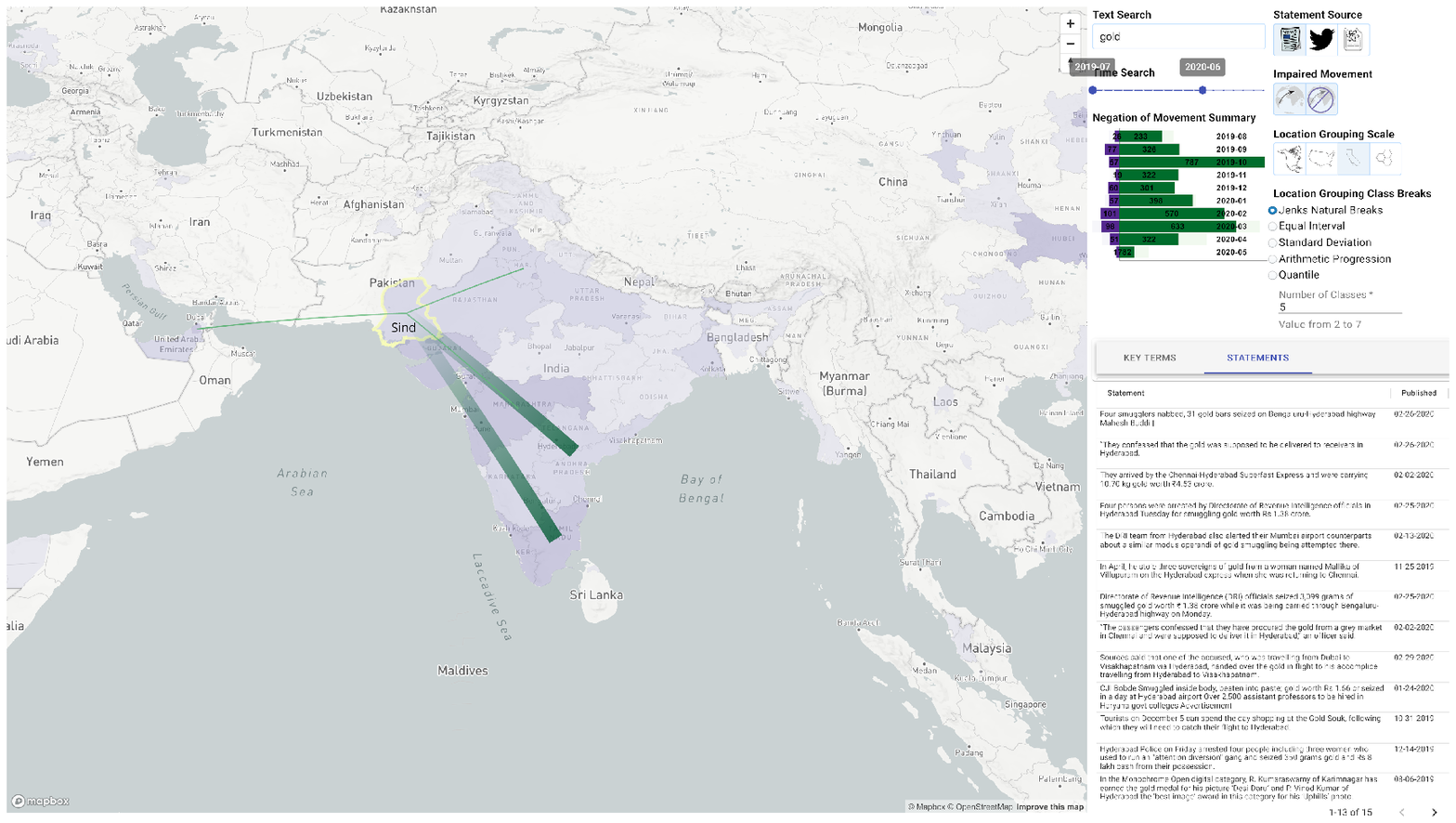}
	\caption{A search for \textit{gold} produces many statements about gold smuggling and possible routes.}
	\label{fig:fig4}
\end{figure*}

Next, after reading some individual statements, she finds out more about multiple cases of gold smuggling from both outside and within India (Figure~\ref{fig:fig4}). To get a more detailed analysis of the movement, she types \textit{gold} in the search box and chooses to aggregate place mentions by the state level. By clicking the neighboring state of Sindh, Pakistan, which has many place mentions in the statements, connections with many states in India are highlighted, including a state in southeast India.

\begin{figure*}[htb]%
	\centering
	\includegraphics[width=\linewidth]{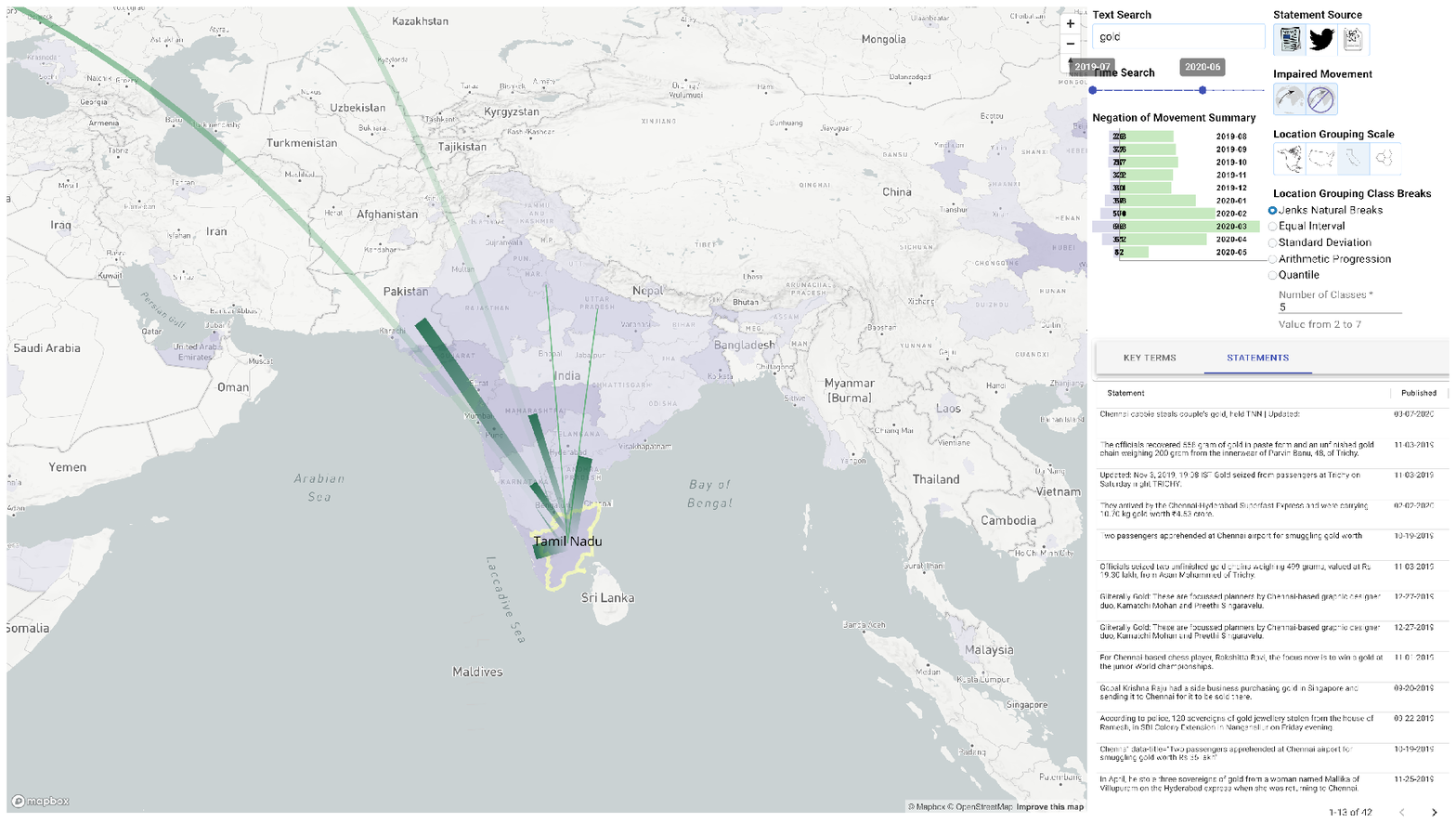}
	\caption{While looking for more detail about gold smuggling, statements suggest a smuggling route from Mumbai to Chennai along the Chennai Express train.}
	\label{fig:fig5}
\end{figure*}

She clicks this state in India and discovers that the state is Tamil Nadu, where the large city of Chennai is located. From reading a few statements, she knows that gold smuggling in Chennai is arriving at the airport and through the Chennai Express train from Mumbai. The connection with Mumbai through the train is confirmed on the map by the strong connection with Mumbai's state, Maharashtra, on the west coast closer to Pakistan (Figure~\ref{fig:fig5}). She can now clearly visualize and report on some of the prominent drugs and gold smuggling sources into and throughout India and read more detailed descriptions about individual incidents.

\subsubsection{Examining the Impact of the Pandemic on Travel for Tourism}

\begin{figure}[htb]%
	\centering
	\includegraphics[width=\linewidth]{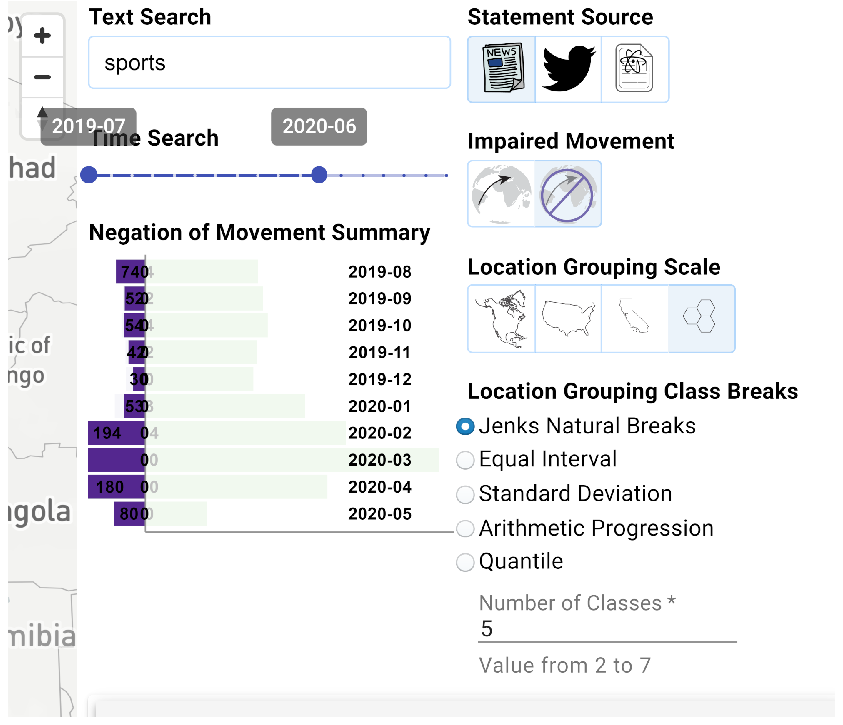}
	\caption{Mousing over the Two-sided Temporal Bar Chart shows the number of filtered statements for impaired movement (purple bars to the left of the axis) involving sports has been increasing since the pandemic began (bars are aggregate counts of statements by month, with the oldest month at the top). Filtered statements by the search are shown in the foreground, while all statements are shown in the background on the mouse hover.}
	\label{fig:fig6}
\end{figure}

Arti Reddy is a travel agent in India. She uses GeoMovement to understand the pandemic's impact on global travel and travel related to India. As of May 2020, like other countries in the World, India was dealing with a global pandemic. Since Arti had previously planned to advertise to potential customers traveling for sporting events, she chose this topic to investigate. She loads GeoMovement and enters the term \textit{sports} in the text search box. With the geospatial hexagon geo-bins as a layer, she quickly sees a hotspot of discussion in Japan. She selects to filter statements for impaired movement. While mousing over the Two-sided Temporal Bar Chart, she sees that the impaired movement statements are more prevalent in recent months (except May, where there is less data), as seen in the Two-sided Temporal Bar Chart in Figure~\ref{fig:fig6}.

\begin{figure*}[htb]%
	\centering
	\includegraphics[width=\linewidth]{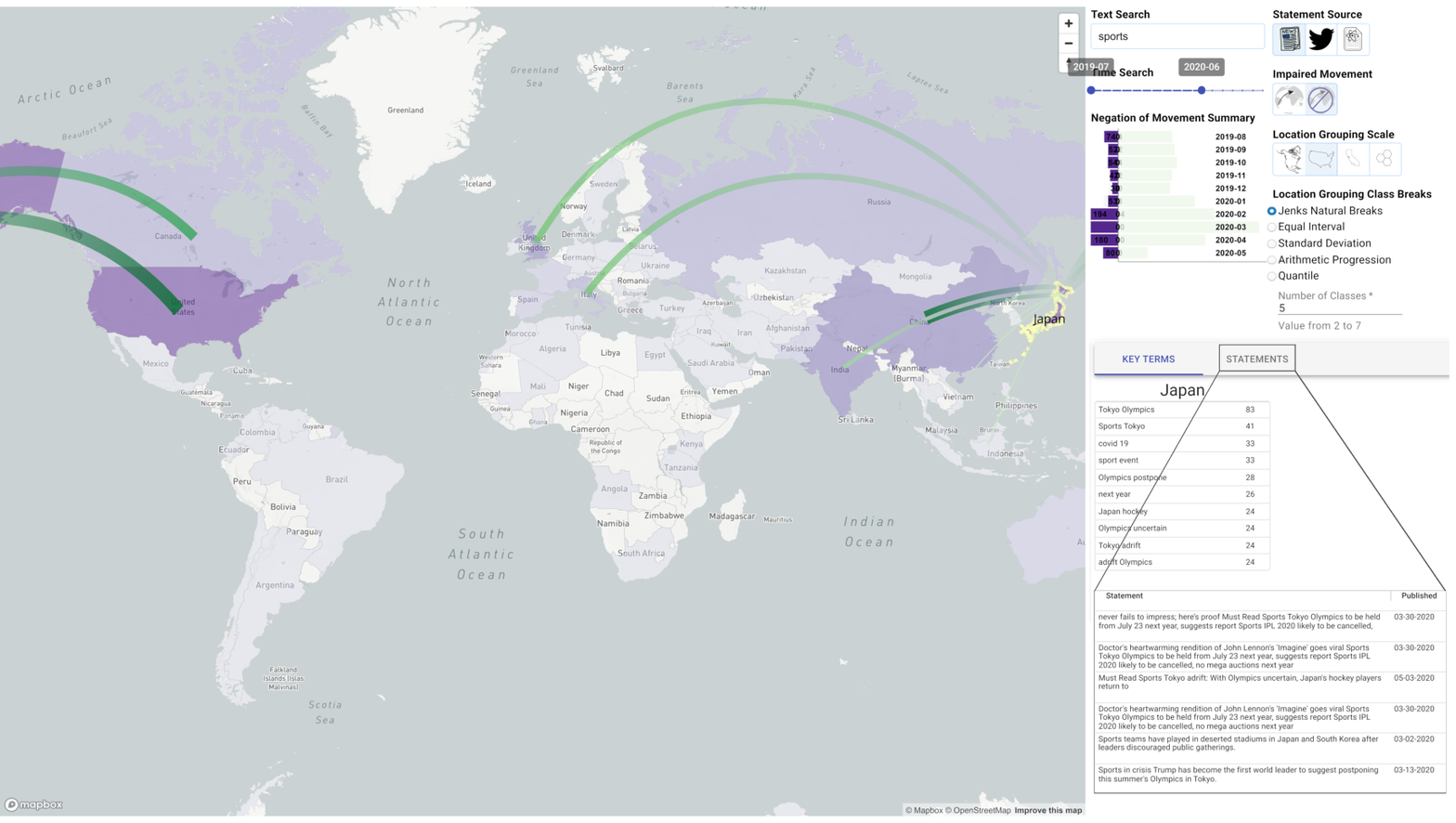}
	\caption{Sports have been disrupted during the pandemic, including the Olympics, where athletes travel from around the World. Mousing over the Two-sided Temporal Bar Chart shows that the filtered statements for impaired movement have increased since the pandemic began.}
	\label{fig:fig7}
\end{figure*}

Arti switches to the country geo-binning, chooses to view both types of movement again, and moves the timeline from July 2019 to December 2019, and Japan is still a popular location. After selecting Japan on the map, she sees in the bi-grams list that the upcoming Olympics in Japan are prominent, and on the map, she sees that there are connections between Japan with many other places in the World (Figure~\ref{fig:fig7}). There is much discussion about the potential impact of the pandemic on the Olympics. She is intrigued to see that there are connections between Japan and her home country that implies her business will be affected.

\begin{figure*}[htb]%
	\centering
	\includegraphics[width=\linewidth]{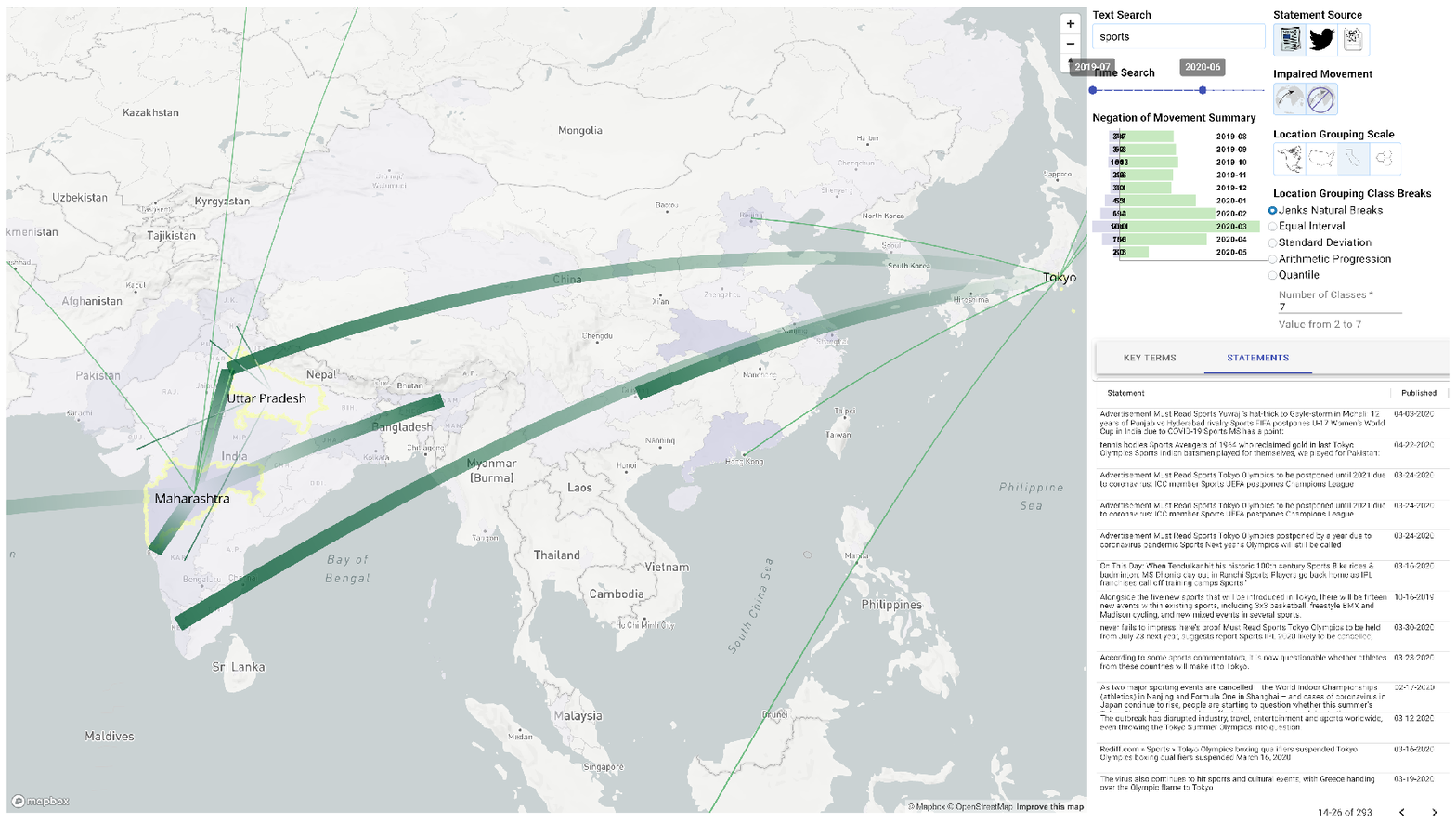}
	\caption{Sports in India have also been affected, including competitions where local teams compete against other teams from around the World.}
	\label{fig:fig8}
\end{figure*}

Since she is most familiar with India's geography, she changes the geo-binning to the state level for more detail and selects two states in India with much discussion (Figure~\ref{fig:fig8}). A quick look through the statements shows that the Olympics are in jeopardy, and a closer-to-home event of the under-17 women's soccer World Cup scheduled to be in India was unfortunately postponed. Teams would have come from around the World for this event. This sad news prompts her to follow up on the story by clicking the statements to view the original articles in her web browser to see if it will be rescheduled and thus if there will be a future need for travel. From reading the statements, she sees that an auto show, the AP World Indoor Sporting Championships, and the Australian Grand Prix auto racing's China leg are three events postponed in China. Cricket in Australia was also severely affected, with teams planning to come from India, England, and many other countries to compete. Major sporting leagues in the U.S., like the NBA basketball league, were also interrupted. The disruptions to nearby events like cricket are particularly concerning given the sport's popularity with Indians and their potential spectator travel related to her business.

\begin{figure*}[htb]%
	\centering
	\includegraphics[width=\linewidth]{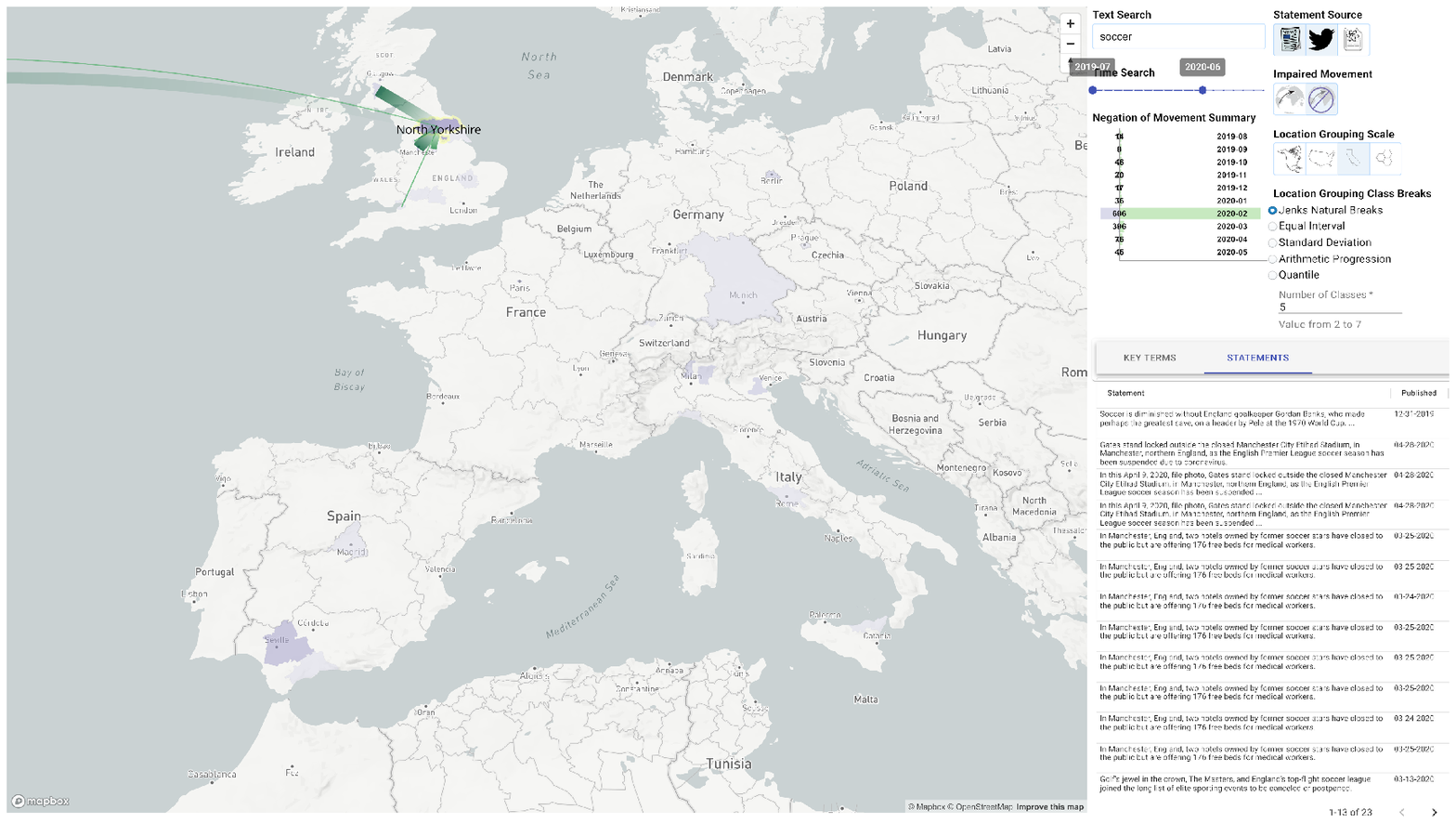}
	\caption{Europe is another example where the pandemic postponed many of their beloved soccer matches.}
	\label{fig:fig9}
\end{figure*}

Lastly, to see how widespread the pandemic's impact is on global sporting events where Indians may travel, Arti explores other countries and sees that the pandemic has significantly impacted professional soccer in Spain and other European countries. Madrid, Milan, and cities in Germany where prominent soccer clubs play all show up clearly on the map as having their games affected (Figure~\ref{fig:fig9}).

These use cases highlight the potential for information foraging and acquisition from statements describing movement. Geovisual analytics that combines multiple views of the data and overview + detail capabilities allow users to quickly identify important locations, connections between locations, time periods, and topics of interest. We showed how the map view could show critical hotspots, and the Two-sided Temporal Bar Chart can show essential time periods, prompting the user to investigate details.

In this section, our use cases show GeoMovement's current sensemaking capabilities. Moreover, the examples of computational prediction errors and suggested solutions highlight how geovisual analytics is well suited for allowing user interactions to either explicitly or implicitly improve GeoMovement's computational predictions. Improved prediction accuracy can make the sensemaking process more efficient by reducing the human effort to sift through incorrect intermediate results.

\subsection{Case Studies Needs Assessment}

In addition to our case studies demonstrating the effectiveness of GeoMovement in extracting information about movement, we used them as a needs assessment to propose additional functionality. In the first case study, Jennifer identified statements from an event that she was not interested in. Although she quickly found what she was looking for, other problems may require much further analysis, and future results should not include statements deemed not relevant. Therefore, a future addition to GeoMovement can consist of a mechanism to either mark statements as completely ``not relevant" and not show them or as ``less important." If user accounts were added to GeoMovement, these preferences in statements could be stored. If certain types of statements will never be relevant to that user, their marked statements can be used to affect their future search results.
Such user input can be used as feedback to the system and in the future, statements that are very similar to those that are marked not of interest will also be filtured out; i.e., the system learns from the feedback.
Elasticsearch allows for on-the-fly criteria provided in searches that promote or demote statements or exclude them, allowing for these criteria to be personalized if user accounts were added and changed for each each search.

In the second case study, Arti found that impaired movement became more prevalent in recent months. It would be a fair assumption that this is because of travel restrictions from the pandemic. However, Arti may ask for more detail about ``How is impaired movement changing over time?" 
To do this, she should easily access statements by month. Currently, she would have to change the time slider filter to each month and select the impaired movement button. A straightforward way to answer this question would be to choose any bar in the Two-sided Temporal Bar Chart to update the time range and impaired movement filters. For example, clicking on the impaired movement purple bar for January 2020 would update all views to show these statements. Similarly, when Jennifer found a spike in smuggling activity in India in October 2009, she would likely ask herself why this is and want a quick way to filter to those statements.

Finally, both Jennifer and Arti were interested in India. A question they both may have is: ``What other locations have similar problems as India?" Currently, GeoMovement allows selection on the map of multiple locations to show bi-grams lists for both locations' statements next to each other. However, there are automated ways to give users hints on locations with similar statements. For example, Elasticsearch provides a ``percolate" query where, after the user already selects a location, they could choose a user interface control and click a second location. The statements from the first location can be used in the percolate query. The result would be statements like the first set, and therefore other places like that place will show.

Our evaluation of the sensemaking capabilities of GeoMovement includes a statistical summary illustrating the challenge in extracting meaningful information without it, case study demonstrations, a needs assessment, and a list of user requirements. We show that despite the large volume of messy data that is at times ambiguous, GeoMovement can quickly extract meaningful information about geographic movement.

\section{Results and Implications}

Analyzing a large volume of text describing geographic movement requires imprecise computational processing and predictions on already messy data that different people can perceive differently, resulting in some errors. We present the data in a geovisual analytics web application that follows the overview-first + detail mantra~\citep{shneiderman1996eyes}. Users can find exciting patterns in the overviews and areas that need further investigation. They can also search the data and eventually drill down to the actual statements and articles. Results that appear to be erroneous or merely uninteresting can be hidden so that it is easier for the analyst to focus on more helpful information.

Our approach tightly integrates geovisual analytics with behind-the-scenes computational predictions for statements describing movement from most text that does not, GIR methods to extract geography for mapping and analysis and a novel prediction for impaired movement. We show the value of GeoMovement through case studies where humans quickly learn about and visualize geographic movements. Furthermore, we demonstrate that a multi-scale system that applies a space-time-concept/attribute filtering process can effectively make sense from large volumes of messy movement statements.

Because of the challenges of utilizing messy big text data for understanding geographic movement, previous work has either focused only on movement data from precise sensors or addressed narrow domains of movement described in text. However, the need for improved methods to utilize big text data about movement is illustrated in work described in Section~\ref{sec:sec2}, and \citet{hultquist2020crowd}, \citet{hultquist2019nuclear}, and \citet{Janowicz2019semantic}, where geography in text complements sensor data or stands alone to solve real-world problems. \citet{maceachren2017bigdata} also encouraged such advances in his positional paper. GeoMovement 1) tightly integrates computational predictions and geovisual analytics, and 2) is built with modern web computing technologies that return results from user queries in milliseconds. These features allow users to make sense of and uncover movement patterns quickly. We show how this efficient and broad movement exploration is unfeasible without computational and visual means to assist a human. Moreover, this completes our research objective to show that modern advances in NLP and IR can leverage large volumes of movement statements to understand the movement described quickly.

Although we stopped collecting text for GeoMovement, our techniques can be adapted in a straightforward technical way to accommodate a continually updating source given modest computational and storage requirements. In addition, all of our code is available with an open-source license to prompt future advances \citep{pez2022geomovement}.

We anticipate GeoMovement or a production version to be beneficial for scientists, journalists, or even the general public to find information about the movement of humans, wildlife, goods, disease, and much more. Advances to analyzing movement statements can provide similar knowledge gains as movement recorded by sensors with precise geospatial and temporal data.

\section{Conclusions and Future Work}

Our research shows that modern computational techniques can be combined with a human-in-the-loop geovisual analytics system to overcome significant challenges and identify, process, and explore large volumes of movement statements to quickly obtain an overview of movement patterns and forage for detailed information of interest. Future research should take advantage of the wealth of context information found alongside geographic movement in text documents about why, when, and how the movement is occurring that is not often present in precise GPS data. Our research methods can likely be adapted to analyze statements involving other attributes like time and more complex spatial analysis like correlation. In addition, future research should explore the integration of precise geospatial movement data with movement in text. One initial way is to link them spatially and temporally. For example, the following steps can involve connecting entities in the text to other information using the Semantic Web \citep{bernerslee2001semantic} and linked geographic data \citep{Stadler2012linked,janowicz2012linked}. Future research needs to identify what makes movement statements different from other statements.  \citep{pezanowski2022movement}, took initial steps towards this goal by identifying vital characteristics of movement statements that humans use to differentiate the movement described.

Also, since we chose to illustrate the effectiveness of GeoMovement through the case studies and other assessments in Section~\ref{sec:assess}, future work should include a more thorough evaluation of user needs, especially focusing on the additions suggested in this Section.

Moreover, prior research shows that a geovisual analytics system combined with ML predictions can continually improve the accuracy of the ML predictions by having users correct machine errors~\citep{snyder2020,andrienko2022va}. An extension of GeoMovement can make it an intermediary between computational predictions and humans. As more humans use GeoMovement, humans can iteratively correct any errors they encounter, thereby improving the predictions. As an example, in Section~\ref{sec:impaired}, we discuss how an ML model would likely be superior to a rules-based approach detecting impaired movement but requires a large amount of training data. GeoMovement can show initial predictions of normal movement and impaired movement from an ML model that used a small amount of training data. A simple tool can allow users to correct errors in the predictions. Once many users are using GeoMovement, the training data set can proliferate. This technique has also shown success in commercial production mapping systems like Google Maps, where users reach their destination, and Maps asks about driving directions' accuracy.

Two overall ways that GeoMovement can improve the computational predictions are, first, explicitly asking for feedback on any incorrect predictions and second, through implicit user actions. For example, in GeoMovement's statements view, a button can be added to explicitly mark any statements that do not describe movement. Once a sufficient number of users mark a statement to conclude that it is incorrect, the corrected statement can be added to the corpus of statements used to train the model. As an example of implicit actions, the statements' list is currently returned in order by the search engine software that roughly corresponds to how closely they match the search terms entered and how frequently those search terms appear in the statement. Skipped statements can be recorded when users page through results to find what they are looking for. If many users skip certain statements, they can be deemed less valuable and given a weighting that lets them be listed lower in the order they are returned, thereby promoting more critical statements. There are many other possibilities to obtain either explicit feedback (users correct place mentions that were assigned to the incorrect location using the GIR techniques~\citep{karimzadeh2019geoannotator}) or implicit feedback from users (identifying important locations to highlight based on where previous users navigated to on the map) to improve GeoMovement's sensemaking ability.

Additionally, none of the existing GIR systems have ideal performance metrics. There is often a trade-off between different geoparsers' false positives and false negatives and other ways to rank geocoding results on large datasets. Future work can allow user controls in GeoMovement to choose different geoparsing back-ends. Also, the open-source GIR GeoTxt can be modified to allow GeoMovement users to set a tolerance to allow more false-positives in situations where missing relevant statements are most critical or allow fewer false-positives in situations where the key is a quick overview and having every relevant statement does not matter. Our use of geovisual analytics for GeoMovement adds substantial potential for future improved analysis of movement statements.

Finally, it is important to revisit that GeoMovement only supports English. Therefore, a valuable extension to our research would be the addition of other languages. The higher-level software, programming languages, and APIs (discussed in Section~\ref{sec:B}) we used in our research to harvest, process, store, retrieve, and visualize data are all capable of handling many languages and character sets. However, incorporating documents in other languages into GeoMovement would require significant work on the three main computational predictions we used. First, identifying statements that describe movement would require a large amount of labeled training data in the added language. Also, for higher accuracy in predictions, language models like ELMo would need to be re-trained for that language like the research of \citet{che2018lang} and \citet{fares2017lang}. Second, our geoparsing methods to extract place mentions and geocode them to their correct location on Earth would need to be adapted \citep{mandl2008lang}. Third, our negation detection techniques would also require adaption to support other languages \citep{morante2021lang}. These changes are possible but need significant work as existing research is less robust than the equivalent in English.

The World is a dynamic place, and understanding geographic-scale movements of things is essential in many domains like business, public health, and environmental science. Leveraging information about movement found in text can complement precise sensor-based movement data. Also, movement described in text is valuable since precise movement data is often unavailable because of high costs or impracticality to deploy sensors. GeoMovement and the combination of computational and visual methods it integrates are steps toward that objective.

\backmatter

\bmhead{Supplementary information}

We include a supplementary video of GeoMovement to show how geovisual analytics helps derive valuable patterns about geographic movement from text documents.

\bmhead{Acknowledgments}

We would like to thank the Information Technology group in the College of Information Sciences and Technology at The Pennsylvania State University for providing computing resources to host GeoMovement. Specifically, Adam McMillen, a Systems Administrator in the group, provided his expertise in establishing a virtual server and deploying GeoMovement to it.

\section*{Statements and Declarations}

No funds, grants, or other support was received.\\

\noindent
The authors have no competing interests to declare that are relevant to the content of this article.\\

\noindent
Author Contributions: Conceptualization: Scott Pezanowski, Prasenjit Mitra; Methodology: Scott Pezanowski, Prasenjit Mitra, Alan M. MacEachren; Software: Scott Pezanowski; Data curation: Scott Pezanowski; Formal analysis: Scott Pezanowski; Validation: Scott Pezanowski, Prasenjit Mitra; Resources: Prasenjit Mitra; Visualization Design: Scott Pezanowski, Alan M. MacEachren; Visualization Implementation: Scott Pezanowski; Project administration: Scott Pezanowski; Writing - original draft: Scott Pezanowski; Writing - review \& editing: Scott Pezanowski, Prasenjit Mitra, Alan M. MacEachren; Supervision: Prasenjit Mitra, Alan M. MacEachren \\

\begin{appendices}

\section{Terms Related to Movement}\label{sec:A}

Table~\ref{tab:verbs} shows the list of verbs related to movement used in our negation detection techniques as retrieved from \url{https://archiewahwah.wordpress.com/2019/04/16/movement-verbs-list/} on January 21, 2022.

\begin{table*}[!htb]
\caption{List of verbs related to movement used in our detection of impaired movement.}\label{tab:verbs}%
\resizebox{\textwidth}{!}{%
\begin{tabular}{ll}
\hline
Advance - move forward.                                    & Pootle {[}informal{]} - proceed casually.                 \\ \hline
Aim (for) - go in the direction of.                        & Pop (in, to, over etc) - quickly visit/pass etc.          \\
Amble - walk casually.                                     & Potter - move in an unhurried way.                        \\
Angle (towards etc) - turn one - s steps towards.          & Pound - proceed with fast, heavy steps.                   \\
Back (out, away etc) - move in reverse.                    & Prance - move flamboyantly, with effected grace.          \\
Barrel - move in a forceful, uncontrolled way.             & Progress - advance.                                       \\
Beetle - hurry like an insect.                             & Proceed - go forward.                                     \\
Belt - move swiftly.                                       & Promenade - take a leisurely walk.                        \\
Bez {[}informal, dialect{]} - zip around.                  & Prowl - move in a shifty or predatory manner.             \\
Bluster - move forcefully yet ungraciously.                & Race - move quickly, in competition.                      \\
Bolt - move swiftly.                                       & Ramble - walk far and wide.                               \\
Bounce - move with elastic motions.                        & Regress - go back.                                        \\
Bound - move quickly with large steps.                     & Return - go back.                                         \\
Bumble - proceed in a clumsy fashion.                      & Roam - proceed with no direction in mind.                 \\
Canter - move fairly, like a horse. &
  Roll - {[}literal{]} proceed in turning motions like a wheel / {[}figurative{]} move steadily. \\
Careen - speed forward uncontrollably                      & Rove - wander far and wide.                               \\
Career - speed forward with little control.                & Run - proceed quickly, both feet leaving the floor.       \\
Charge - move aggressively towards something.              & Rush - move with haste.                                   \\
Crawl - {[}literal{]} go on all fours / {[}figurative{]} proceed slowly. &
  Sashay - move in a confident and flamboyant way. \\
Creep - move sneakily or slowly.                           & Saunter - walk arrogantly, confidently.                   \\
Dance - move rhythmically.                                 & Scamper - run like an agitated animal.                    \\
Dart - go swiftly.                                         & Scarper - run away.                                       \\
Dash - run quickly.                                        & Scoot - proceed at a fair pace / shuffle to one side.     \\
Dawdle - proceed slowly and reluctantly.                   & Scud - move quickly as if blown by the wind.              \\
Dive - descend quickly.                                    & Scuff - walk in a careless, friction-producing way.       \\
Dodder - move unsteadily, as if elderly.                   & Scurry - hurry like a small animal.                       \\
Dogtrot - move at a brisk, comfortable pace, like a dog.   & Scuttle - hurry like an insect.                           \\
Emerge (from) - come out of.                               & Seethe - proceed like oozing liquid.                      \\
Escape - move out of danger/confinement.                   & Shuffle - walk slowly, without lifting one - s feet.      \\
File (in) - {[}of multiple people{]} go one-by-one.        & Skedaddle - depart in haste.                              \\
Flee - run away from.                                      & Skip - proceed bouncing from one foot to the other.       \\
Flounce - move in a flamboyant way.                        & Skitter - move hurriedly.                                 \\
Flop - move loosely.                                       & Slide - move frictionlessly.                              \\
Fly - {[}literal{]} move through the air / {[}figurative{]} proceed swiftly. &
  Slink - go smoothy/sensuously. \\
Footslog - march a long distance.                          & Slip - move frictionlessly / make an accidental movement. \\
Forge (on, ahead) - proceed strongly and steadily.         & Slither - slide forward like a snake.                     \\
Gallop - move quickly, like a horse.                       & Slope - as sneak.                                         \\
Gambol - proceed in a playful, energetic manner.           & Sneak - Proceed surreptitiously.                          \\
Glide - move frictionlessly.                               & Speed - move very fast.                                   \\
Go - basic movement verb / depart.                         & Split - depart.                                           \\
Hare - proceed extremely quickly, like the animal.         & Sprint - run at top speed.                                \\
Hasten - move with haste.                                  & Stagger - move unbalanced, unsteadily.                    \\
Head (towards, for, to etc) - proceed in the direction of. & Stalk - move as though hunting.                           \\
Hie {[}archaic{]} - go quickly. &
  Stampede {[}multiple people{]} - progess chaotically /in agitation. \\
Hightail {[}informal{]} - move quickly.                    & Steam - power forward.                                    \\
Hike - go a long distance.                                 & Step - move with the feet.                                \\
Hop - {[}literal{]} proceed on one foot / {[}figurative{]} make a short journey &
  Streak - move quickly, as if leaving a line of light behind you. \\
Hurtle - move quickly, violently and recklessly.           & Stride - walk purposefully.                               \\
Issue (from) - come out of.                                & Stroll - walk in a brisk, leisurely manner.               \\
Jog - move at a medium pace/half-run.                      & Strut - walk stiffly / arrogantly.                        \\
Jump - propel oneself through the air.                     & Swagger - move arrogantly.                                \\
Jaunt - go on a short trip.                                & Sweep - proceed swiftly.                                  \\
Journey - travel a distance.                               & Tank - progress swiftly and forcefully.                   \\
Labour - move with difficulty, requiring force.            & Tiptoe - proceed lightly, silently on the toes.           \\
Leap - jump far.                                           & Traipse - walk a distance.                                \\
Leg (it) - run (away).                                     & Tramp - walk a distance.                                  \\
Limp - proceed unevenly / with an injured leg.             & Trample - walk without precision or care.                 \\
Lollop - proceed in ungainly bounds.                       & Travel - move a distance.                                 \\
Lope - move in large strides.                              & Tread - move using the feet.                              \\
Lunge - jump forward to attack.                            & Trek - travel a long time / distance.                     \\
March - move steadily/forcefully/with purpose.             & Trip - proceed lightly, gaily.                            \\
Meander - proceed in an indirect way.                      & Tromp - walk heavy-footed.                                \\
Mooch {[}informal{]} - go around in a skulking manner.     & Troop - march with effort.                                \\
Mosey - walk in a leisurely manner.                        & Trot - move briskly like a horse.                         \\
Move - basic verb of movement.                             & Trundle - move arduously like a cart.                     \\
Nip (into, across, over etc) - quickly go.                 & Tumble - fall / spiral forward                            \\
Pace - walk steadily.                                      & Undulate - proceed in wavy motions                        \\
Pad - walk casually/softly/steadily like an animal. &
  Waddle - Walk in ungainly fashion, from side to side. \\
Parade - proceed in an extrovert manner.                   & Walk - go by foot.                                        \\
Patrol - walk around in order to guard.                    & Wander - travel without a direction in mind.              \\
Patter - go with a light tripping sound.                   & Wend - travel by a circuitous route.                      \\
Pass - move beyond.                                        & Whizz - go speedily.                                      \\
Pelt - move quickly, like a hurled stone.                  & Wobble - move unsteadily.                                 \\
Perambulate {[}formal, rare{]} - walk.                     & Zip - move swiftly.                                       \\
Plod - move with heavy, laborious motions.                 &                                                           \\ \hline
\end{tabular}%
}
\end{table*}

Table~\ref{tab:adjs} shows the list of adjectives related to movement used in our negation detection techniques as retrieved from \url{https://dspace.ut.ee/bitstream/handle/10062/18053/adjectives\_and\_adverbs\_of\_movement.html} on January 21, 2022.

\begin{table*}[!htb]
\caption{List of adjectives related to movement used in our detection of impaired movement.}\label{tab:adjs}%
\resizebox{\textwidth}{!}{%
\begin{tabular}{l}
\hline
In the text, some nouns are qualified with adjectives (sharp, sudden) and some verbs with adverbs (dramatically, significantly).\\
Adjectives used to describe movement include: slow, slight, moderate, gradual, steady, quick, rapid, significant, sharp, substantial, dramatic. \\
Used to show a small change: slight \\
Used to show a regular movement: gradual, steady \\
Used to show considerable, striking or unexpected change: significant, substantial, dramatic (both positive and negative change), sharp, sudden \\
Adverbs are formed by adding -ly to the adjective, and sometimes one or two other letters change as well. \\
Degree of change: \\
dramatically, considerably, significantly, substantially, sharply, moderately, slightly \\
Note that ``dramatically" can refer to both good and bad changes. \\
Speed of change: \\
rapidly, quickly, suddenly, gradually, steadily, slowly
          \\ \hline
\end{tabular}%
}
\end{table*}
\section{Detailed Description of GeoMovement}\label{sec:B}

\subsection{Text Data Sources}\label{sec:textsources}

First, we purchased news articles from Webhose.io (\url{https://webhose.io/}). Webhose.io collects text documents from various Web sources and distributes them in a clean semi-structured XML format. To save costs and reduce the number of irrelevant statements, we used their web archive tool to obtain text documents that match the key term \textit{travel}, that is in the \textit{English} language, and that was published on a website categorized as \textit{news}. We downloaded all articles in their archive between August 1, 2019, and May 10, 2020, that fit these filters (397,919 articles).

Second, we acquired and incorporated tweets using the GeoCov19 dataset spanning February 1, 2020, to May 2, 2020~\citep{qazi2020}. This dataset contains hundreds of millions of tweets related to the Covid-19 pandemic. We harvested only the tweets in English, which resulted in 328 million tweets.

The third source for statements is the Covid-19 Open Research Dataset (CORD-19)~\citep{wang2020cord}. This dataset contains scientific articles about the Covid-19 pandemic that were harvested and placed in a clean semi-structured XML format. The collection has over 50 thousand articles retrieved from PubMed Central (\url{https://www.ncbi.nlm.nih.gov/pmc/}), bioRxiv (\url{https://www.biorxiv.org/}), and medRxiv (\url{https://www.medrxiv.org/}) that match search terms related to all Coronaviruses and is updated daily. We downloaded the dataset on November 14, 2020, and selected documents from that date back until August 1, 2019, which totaled about 15,600 documents.

\subsection{Application Development for Efficient Sensemaking}\label{sec:B4}

GeoMovement is an entirely web-based application. We built the web client using the React JavaScript Library (\url{https://reactjs.org/}) as its framework. We also used Material UI (\url{https://material-ui.com/})  React Components to make development efficient and provide a friendly and familiar user experience. We used Mapbox GL JS (\url{https://docs.mapbox.com/mapbox-gl-js/api/}) in the map portion of GeoMovement and Deck GL Libraries (\url{https://deck.gl/})~\citep{Wang2017deck} to provide advanced mapping and geovisualization capabilities along with state-of-the-art rendering speeds for large amounts of data. The web client application is a visual interface for humans to search the server's data in Elasticsearch. Other software used to pre-process the data includes Postgresql (\url{https://www.postgresql.org/}) for data storage and PostGIS (\url{http://postgis.net/}) for geospatial computations. The result of this application design is a web application accessible in any major modern web browser, both desktop, and mobile that produces rapid results to user searches of hundreds of millions of descriptions of geographic movement. Almost instantaneous responses to searches enable users to more easily forage through data in the sensemaking process and identify important information. Figure~\ref{fig:fig1} shows the user interface of GeoMovement.

\section{GeoMovement User Requirements}\label{sec:C}

In Section~\ref{sec:cases}, we address the utility of GeoMovement through two case study scenarios that demonstrate the ability of GeoMovement to support information foraging through the large volumes of text coming from multiple kinds of sources. The scenarios make specific assumptions about the required capabilities, skills, and prior knowledge of users.

\begin{center}
	User Capabilities
\end{center}

Currently, GeoMovement works only with text in English; thus, users are expected to be English speakers. Given the focus on interactive maps and graphical displays, users need manual dexterity sufficient to handle interaction with the map and various controls and adequate visual acuity to read the maps and graphs.
Although we did not test GeoMovement on many computer configurations, based on our internal testing, web development expertise, and established guidelines and recommendations for the development technologies, we can estimate that the user needs modest computing capabilities. The user should have a computer with a minimum monitor screen resolution of $1024 \times 768$ pixels. A modern JavaScript-enabled browser (Chrome, Firefox, Safari, and Microsoft Edge) is required, and minimum computer hardware specifications to support these browsers. GeoMovement was tested and worked on a modern smartphone with a mobile Chrome browser; however, GeoMovement's sensemaking capabilities are restricted because of the small touch screen. Although GeoMovement's design and architecture provide swift search responses from the server software, performance depends on the users' Internet connection speed because it is web-based. Therefore, GeoMovement will function, but user Internet connection speeds under the typical broadband minimum of 25 Mbps will likely produce a lag in user search results that inhibits the analysis.

\begin{center}
	User Skills
\end{center}

GeoMovement expects familiarity with web browsers, graphs, and maps. We assume some experience with thematic maps (maps that depict data geographically), but no particular expertise is required. Thus, users familiar with the kinds of maps routinely found on news sites like those of the New York Times of the Guardian are sufficient. An ability to understand simple statistical graphs (bar charts) is assumed.

\begin{center}
	User prior knowledge
\end{center}

Since GeoMovement is designed for user-led information foraging, it does not provide suggestions to users about what to explore or specifically what to look for. The assumption made is that each user has some knowledge of the topic they are interested in, that they can apply, to leverage the visual interface and computational methods included in GeoMovement. That is, in fact, the point. This is a human-in-the-loop system intended for users who have domain expertise that is complemented by the ability of GeoMovement to process very large volumes of text and potential for interactive filtering tools to support quick narrowing in on items of interest through the use of that domain knowledge to decide what is interesting and what is not interesting. Beyond domain knowledge, however, users also need some understanding of the limitations of the computational methods that underlie GeoMovement. Specifically, they should understand that any system like this will generate some false positives and some false negatives; thus, it will show them some irrelevant information (which their domain expertise is likely to allow them to ignore) and miss some relevant information. Therefore, they need to understand that absence of evidence from GeoMovement does not necessarily mean the absence of the phenomena they are interested in.




\end{appendices}


\bibliography{ExploringDescriptionsOfMovementThroughGeovisualAnalytics}


\end{document}